# Trinity by the Numbers: The Computing Effort that Made Trinity Possible

Nicholas Lewis
Los Alamos National Laboratory
Los Alamos, NM 87545

**Abstract:** This article addresses shortcomings in the existing secondary literature describing the nature and involvement of computing at the World War II Los Alamos Lab. Utilizing rarely used source materials, and identifying points of bias among more commonly used sources, this article provides a more complete representation of wartime Los Alamos' computing operations and personnel, including the Lab's typically under-represented human computers, and how they contributed to the success of the Trinity test. This article also identifies how the Lab's unusual wartime computing demands served as a formative experience among many Los Alamos personnel and consultants, who contributed significantly to the development and use of mechanized computing at and beyond Los Alamos after the war.

## Introduction

The Trinity test marked the spectacular inauguration of the atomic era, and it completed the primary objective of the largest scientific research and development effort in history. The test also validated one of the most extensive computational efforts assembled up to that time, which served as a vital component of the Manhattan Project's success. Many histories of the Manhattan Project have examined the scientific effort at Los Alamos and its work toward the success of the Trinity test. The importance of computing at Los Alamos toward the success of Trinity has received far less attention in the existing literature. The fragmentary examinations that do exist feature notable gaps and biases that have perpetuated an incomplete representation of many aspects of computing at Manhattan Project Los Alamos. This article draws from a variety of sources, both frequently and rarely used, to provide a more thorough representation of how the complex technical and human networks responsible for the Lab's computing capabilities converged at wartime Los Alamos, and to demonstrate how they were an essential component for the success of the Trinity test. Those networks, through the transmission of key concepts and the influence of well-placed individuals, along with the increasing demands they placed on computer development at and beyond Los Alamos, also served as a catalyst in the rapid advancement of computing technologies and methods following the Second World War.[1-3]

## Origins of Manhattan Project Computing

The first organized computing effort of what became the Manhattan Project began at Berkeley in the Spring of 1942. Taking place nearly a year before the founding of the laboratory at Los Alamos, the Berkeley computation work contributed to the ultimate decision to build a dedicated weapons-development lab, and to the subsequent computing effort involved with the success of that lab's mission.

Through the end of 1941, the US atomic-bomb program, then under the auspices of the Office of Scientific Research and Development (OSRD), had focused primarily on slow-neutron fission research. The purpose of this focus was to develop techniques for generating the needed isotopes of uranium and plutonium in sufficient quantities for an atomic weapon. Without a reasonable confidence in the project's ability to produce the needed fissile materials, there would have been little purpose in developing the weapon itself. In late 1941, Arthur Compton, the Nobel laureate and head of the Chicago Metallurgical Laboratory, provided encouraging progress reports to the OSRD about the multiple methods being pursued to produce fissile materials. He also provided critical-mass estimates that pointed toward the feasibility of a weapon. The reports resulted in an increased project focus on the issues of atomic-bomb design and construction. Engineering the weapon itself meant expanding on the relatively small amount of research on fast neutrons—neutrons with high kinetic energies needed to drive the detonation of an atomic weapon—conducted up to that point.[4-7]

Although responsible for the physics of weapon development, the Chicago Met Lab's internal focus on a slow-neutron chain reaction to produce plutonium had left most of the project's fast-neutron research in the hands of subcontractors. The Berkeley Radiation Lab under Ernest Lawrence was among those subcontractors, housing Lawrence's calutron project for electromagnetically separating isotopes of uranium. J. Robert Oppenheimer, a Berkeley theoretical physicist, first became involved with the atomic-bomb effort in October of 1941 when Lawrence invited him to join the Radiation Lab's uranium separation work.[8, 4, 6]

In December of 1941, just after Pearl Harbor, Oppenheimer asked theoretical physicist Robert Serber to come to Berkeley from his faculty position at the University of Illinois, Urbana. The site of his postdoc, Serber was to help Berkeley prepare for a theoretical-physics conference that Oppenheimer would be hosting there in the summer of 1942. Oppenheimer had become closely tied to the small but increasing amount of fast-neutron research taking place among the Met Lab subcontractors, and the conference of "luminaries" was intended to discuss the theory of a fast-neutron reaction, and the possible design of an atomic weapon.[4, 6]

Serber, in connection with his research at Berkeley on explosion hydrodynamics—broadly, the flow of physical materials—and the theory of efficiency, recruited two of Oppenheimer's stand-out postdocs, Stanley Frankel and Eldred Nelson. The two had previously conducted magnetic field orbit calculations for the calutron project, and Serber enlisted their computational talents to work on the problem of neutron-diffusion theory, a description of how neutrons move and distribute themselves in a mass.[8]

Little or no neutron-diffusion research had taken place in the United States up to the sharing of the 1941 MAUD Report from the British atomic-bomb program, which provided research and encouraging early results about the feasibility of a fission weapon. Gregory Breit, the University of Wisconsin physicist initially in charge of the OSRD's fast-neutron research, had not conducted any known work on neutron diffusion before he resigned from the position in May of 1942, citing slow progress and security concerns. Arthur Compton quickly replaced Breit with Oppenheimer, who had earlier that year assisted Compton in estimating weapon efficiencies—the amount of energy released from the weapon on detonation, compared with the total amount of energy available.[9, 8, 6, 7]

With only a sparse literature to guide them, Frankel and Nelson did more than improve upon the simple diffusion theory the British had provided. According to Serber, the two theoretical-physics postdocs created an exact integral equation that more adequately described the complex neutron diffusion that would occur within a weapon. As a consequence, having assembled a hand-computing group equipped with mechanical desktop calculators, Frankel and Nelson were able to perform more accurate diffusion-theory calculations on the critical mass of a weapon than had previously been possible, and they were able to estimate its efficiency. Estimates of critical mass—the minimum mass of fissile material needed to maintain a fission chain reaction—had varied considerably over the preceding year, largely due to the limited neutron-diffusion research up to that point. Sound estimates of how much material would be needed to build a weapon, and of the effectiveness of that weapon, were essential for drafting an actionable conference report for the OSRD.[7, 8, 4]

Recommendations from the Berkeley report, in conjunction with reports from experimentalist conferences in Chicago, would inform the decisions of the project's leadership on the logistics of fissile-material production, the weapon-development process, and the overriding concern of whether the United States should commit to completing the atomic-bomb effort. Of particular importance for the continuation of the atomic-bomb program, the Berkeley conference offered key input from the theoretical physicists on an atomic device's feasibility as a weapon.[8, 4]

Serber, Frankel, and Nelson were the first to present at the conference. Their calculations made them confident that a gun-type weapon—which would fire two, sub-critical masses together to form a single critical mass—was feasible to construct, but concluded that such a weapon would require up to six times the critical mass of some previous estimates. With the Berkeley group's preparatory work, the other luminaries at the conference considered the A-bomb to be a "settled issue" only two days into the event, leaving room for Edward Teller to distract the attendees with his idea for a "Super," a thermonuclear weapon. Despite Teller's distractions, the group concluded in its report that, while feasible, a significant scientific and technical effort still lay ahead to build a working atomic device.[8, 4]

After receiving the Berkeley group's conclusions, the OSRD committee responsible for uranium research recommended that the US commit to developing an atomic bomb, arguing that such a weapon could win the war. Following approval from Franklin Roosevelt, the project moved forward.[4]

**Project Y and T Division**

In the summer of 1942, the US Army formally established the Manhattan Project and began to transfer responsibility for atomic-bomb development from the OSRD. By that autumn, citing the slow progress of the previous year, the formal decision to construct an atomic weapon also recognized the need to centralize the ordnance work to complete the weapon as quickly as possible. The US Army general and newly minted director of the Manhattan Project, Leslie Groves, followed Arthur Compton's suggestion to select Oppenheimer as the head of Project Y, the as-yet unbuilt Lab intended to design and construct the atomic weapons. Oppenheimer suggested a location for the Lab based on his experiences in the mountains of Northern New Mexico.[4]

Following their success at Berkeley, Oppenheimer recruited Stanley Frankel and Eldred Nelson to come to the new Lab, where they would continue their work on neutron diffusion for Los Alamos' primary objective, the gun-type weapon. At that point, either uranium or plutonium were to fuel the weapon. A precursor of the implosion concept—where a subcritical amount of fissionable material would be compressed radially into a supercritical state—was first proposed at the Berkeley summer conference, but received little attention at the time, as the conference luminaries believed the gun-type concept to be the more straight-forward design choice.[7, 10, 8]

Among the first to arrive at Los Alamos in March of 1943, Frankel and Nelson were assigned to the Theoretical or "T" Division, which played a crucial role in the Lab's weapons-development effort. Early on at Los Alamos, T-Division staff were responsible for critical mass and efficiency estimates needed to guide the bomb's specifications and physical design. As the weapon designs matured, T Division was pressured to provide more exacting specifications for the groups ordering or fabricating the weapon components. T-Division estimates were also essential to continue planning uranium enrichment and plutonium production, the complexities of which would have significant weapon-design implications for Los Alamos. Later in the project effort, T Division had to model the behavior of the proposed weapons, including the damage they would potentially cause. For an in-depth discussion of T Division's staff and social history, see J. Shlachter's article in this issue.[12] A key component of T Division's estimates, Frankel and Nelson were under pressure to continue their refinement of neutron-diffusion theory. A better understanding of how neutrons would diffuse during the assembly of the gun-type weapon would help Lab theoretical physicists determine the possibility of the chain reaction starting prematurely, which would reduce the effectiveness of the bomb. To complicate matters, if the bomb designers opted to use a cylinder of fissile material in a weapon, which would be easier to fabricate than a sphere, T Division needed to know how much of the yield might be lost due to that concession. However, determining the behavior of the "odd-shaped bodies" that would occur as a cylinder underwent assembly—the process of forming a critical mass—was far more difficult than for a sphere.[8, 6, 11, 4]

These problems were not solvable analytically, and because of time constraints and the lack of expensive fissionable materials available for testing theory, T Division had to rely heavily on numerical methods to supply its critical mass and efficiency estimates. A great deal was unknown early on at the Lab about the properties of the nuclear materials being studied. Although, theory and experiment had improved the Lab's understanding of uranium 235 far more than plutonium, which had only been discovered in 1940. Plutonium intrigued project scientists, as it seemed possible to produce plutonium more quickly than uranium 235 could be enriched, and it might have been needed in smaller quantities in a weapon. The limited information available meant that T-Division staff had to work with experimentalists to obtain whatever measurements were possible to refine theory. The relationship went in both directions, as T-Division scientists aided experimental groups in interpreting the results of their experiments through theory, often needing the aid of the hand-computing group that Frankel and Nelson established after their arrival.[11, 4, 6]

Dependent as they were on computation, T-Division scientists had to balance the detail and accuracy of their calculations with speed. Meeting the wartime pace meant approximating and simplifying the complex variables in a calculation as needed to avoid consuming undue time and resources. Historians Lillian Hoddeson and her co-authors of *Critical Assembly* contended that the ability of T-Division scientists to create meaningful approximations of the problems at hand was a key component in the swift completion of the Lab's mission.[5, 10, 6]

The significance of Los Alamos theoretical physicists developing this computational acumen amid the Lab's frenetic wartime pace should not be underestimated. Theorists had traditionally avoided computationally demanding areas of research before the 1940s, owing to the minimal computing capabilities and the other limited resources typically available to scientists before the war. The Manhattan Project's demands on theoretical physicists, and its government largesse, introduced many theoreticians to the utility of organized computation for solving otherwise intractable problems. This first-hand experience would have formative consequences for scientific research at Los Alamos and elsewhere in the coming decades, as theoretical physicists and other scientists recognized the power of computing to push beyond the traditional research scope of their respective fields. This sea change in the relationship between computing and the sciences at Los Alamos did not begin with large-scale machines, but with far more modest equipment, and calculating methods that dated back centuries.[13, 14, 10]

**The First Computers**

Much of the Lab's wartime mission was predicated on T Division's ability to set up and perform reasonably accurate and actionable computations in very little time, initially relying on minimal information about the phenomena being studied. Facing these daunting pressures, Frankel and Nelson assembled the first computing resources at Los Alamos. Based on their experiences at Berkeley, the two ordered a collection of Marchant, Friden, and Monroe electromechanical desktop calculators. Some of the machines were distributed to project scientists, while others were allocated to T Division's hand-computing operation, which Frankel and Nelson assembled to address the neutron-diffusion problem, and to provide the Lab with a central computing pool. The computers who provided this key resource at Los Alamos were part of a centuries-old tradition in mathematics and the sciences.[11, 10]

A "computer" was originally a person who performed calculations by hand, with organized groups of human computers first appearing in the eighteenth century, usually in support of creating astronomical or navigation tables. By the early twentieth century, astronomy and meteorology were among the largest users of human computing groups. As women began to enter the workforce in larger numbers, the demographics of computing changed from being mostly men in the early-to-mid nineteenth century, to being mostly women in the twentieth. At Los Alamos and elsewhere, women working as computers often transitioned into programming and operating electronic computers as they were introduced between the 1940s and 50s.[15]

Unlike other Los Alamos divisions, T Division was originally organized around projects rather than groups. The hand-computing operation, later designated as T-5 Computations when T Division formally established groups in early 1944, was assembled soon after the founding of the Lab. At first, the computers at Los Alamos were mostly women drawn from townsite volunteers. Leslie Groves considered the support of civilians on the Hill, the common nickname for Los Alamos, to be a waste of resources. Because they required no additional housing or clearance investigations, Groves encouraged the adult family of the scientific staff to work at least part-time. Later, some of the civilian computers were recruited from beyond the townsite.[5, 16, 17, 6]

Despite its importance to the Lab's mission, the hand-computing effort at Los Alamos typically receives only cursory mention in historical accounts, where it is usually presented as a mere precursor to the mechanized operation (which will feature later in this article). Hand computing disappears altogether from most historical discussions after the IBM equipment is introduced into the narrative. These treatments tend to obscure the role of hand computing as an essential Lab resource, which evolved and operated throughout the war, and continued to operate at Los Alamos until the mid-1950s. This common omission is partly due to the first and most-cited accounts of the Lab's computing history originating with figures whose experiences and interests were primarily in machine-based computing. Electronic-computing advocates, such as Nicholas Metropolis, Herman Goldstine, and others, tended to dominate the early history-of-computing narrative, resulting in hand computing being depicted as more of a curiosity than as the most common and widely utilized form of computing through at least the 1940s. Particular to Los Alamos, as a central resource, the human computers worked on a variety of problems across the Lab's wartime effort, compared with the mechanized operation, which was focused almost entirely on a single problem. This made the contributions of the hand-computing group more difficult to identify, and contributed to its underrepresentation in Lab computing history.[17]

The following section draws upon infrequently used sources to integrate better the hand-computing operation and the experiences of the wartime human computers into the larger narrative of Lab computing. Most of what is known about the human computers themselves at Los Alamos is through oral-history interviews conducted long after the fact. Much of what follows is based on interviews the historians Ruth Howes and Caroline Herzenberg conducted in the 1990s for their book, *Their Day in the Sun: Women of the Manhattan Project*, in addition to earlier interviews conducted by the Los Alamos Historical Society, the historian Martin Sherwin, and Los Alamos Computing Division retiree William "Jack" Worlton.

In the Spring of 1943, T Division's newly formed hand-computing operation consisted of about a dozen volunteers, many of whom were spouses of Lab scientific staff, such as Augusta "Mici" Teller, Betty Inglis, Jean Bacher, Kathleen "Kay" Manley, and Beatrice Langer. Although, as the demand for computing intensified, some of the later civilian recruits for the hand-computing operation came from the nearby communities of Española and Santa Fe, and even from outside New Mexico. Josephine Powers, for example, came to the Lab in September of 1943, having worked as a computer for the Department of Agriculture, before applying by correspondence for a computing job at Los Alamos—although the Lab's name and exact location would have been unknown to her when she applied. Another computer, Margaret Johnson, who joined Los Alamos later in the war, was among the civilian computers hired from the local area. Originally from Española, Johnson earned a bachelor's degree in mathematics from the University of New Mexico, and entered graduate school before going to work to support the war effort. Stopping unannounced at the Manhattan Project's Santa Fe office, Johnson asked if the project needed someone with her training. Johnson recalled that everyone in the area knew about the "secret" government project, simply not what it was working on. She was hired into T Division three weeks later.[16]

Working as a computer at the Lab and operating the desktop calculators required some mathematical knowledge. Many of the early volunteers had degrees in mathematics and the sciences, or other relevant training that made for an easy transition. Kay Manley, for example, who was married to the physicist and principal aide to Oppenheimer, John Manley, had undergraduate degrees in mathematics and English. Beatrice Langer, whose physicist husband, Lawrence, famously slept atop the Little Boy bomb, had extensive informal mathematical training from Indiana University. Each began computing work almost immediately after their arrival. The chemistry and ballistics expert Joe Hirschfelder, whose office was adjacent to the computing operation, conducted computer training courses to bring volunteers with less experience in mathematics up to speed.[16, 18, 10, 17]

For families with two parents working on the project, the Lab provided a nursery school and offered improved cleaning services over what was generally available. Some volunteers joined the technical staff for access to these benefits. Kay Manley credited the availability of childcare for the Manley's two children with her ability to work for T Division. Flexibility to plan work schedules around childcare needs also made it easier to recruit computing personnel. Jean Bacher and Mici Teller, for example, alternated working mornings and afternoons in order to trade childcare responsibilities around the nursery school schedule.[19, 20, 16]

Although Stanley Frankel and Eldred Nelson launched the computing operation, Mary Frankel, married to Stanley, stepped in as its first supervisor. Joining the Lab as a junior scientist, and with degrees in psychology and mathematics, Mary Frankel became an expert in numerical methods in physics, and drafted the problems distributed to the computers. Sometimes as much as twenty years younger than the computers she supervised, Frankel expected excellence from her staff, not holding back praise or critique of their performance. Frankel encouraged her staff to study mathematics in their off hours, and suggested those who showed the most promise to build careers for themselves in computation.[16]

Most of the problems sent to the hand-computing operation required between a part of a single day and a few days to process. About five or six computers sitting at tables with desk calculators worked over a four-hour shift, performing calculations based on the instructions Mary Frankel wrote out on a specialized form. When a scientist submitted a problem to the group for calculation, for example, the analysis of nuclear cross section measurements, or the reduction of experimental data, the problem would be broken down into simpler, discrete components. That way, multiple computers could work on the calculation simultaneously, in an early form of parallel processing. This improved calculating speeds and reduced the likelihood of error over having a single computer process the entire problem from start to finish. The computers were also instructed on the needed decimal-digit precision of each calculation, avoiding inadequate precision with too few digits, or wasting time with too many. The calculations from each computer were then combined with the others to provide the complete result.[20, 10, 17, 21]

Located on the second floor of a narrow, hastily constructed wood-frame building, the hand-computing group was situated between Ashley Pond—the centerpiece of the Lab's primary technical area—and the unpaved main thoroughfare, later named Trinity Drive. In the shared office space, the desktop calculators produced a tremendous amount of noise, sometimes making concentration difficult and increasing the likelihood of errors. The building's lack of air conditioning also meant that, during the warmer months, dust from heavy truck traffic entered through the open windows and caused excess wear and tear on the calculators. To safeguard against entry errors and malfunctioning equipment, the group ran the more significant calculations twice, along with Mary Frankel or another supervisor conducting intermediate check points.[16, 17, 10, 22]

If a military-inspired "drop test" failed to correct a broken calculator, it initially went to Santa Fe or Albuquerque for repair. T-Division physicists Nicholas Metropolis and Richard Feynman, the latter of whom led the division's work on diffusion problems, regularly interacted with the computing operation for equations-of-state calculations. After discovering that sending machines off-site delayed needed computations, the two assembled an ad hoc repair service out of their shared office, which reduced the backlog of broken machines until Army recruits took over onsite repairs.[21, 23, 4, 11]

At the end of the first summer, the hand-computing group expanded to about twenty personnel, largely through the addition of human computers from the Women's Army Corps (WAC), the women's branch of the US Army, which converted to active-duty status in July of 1943. Although paid less, computers from the WAC worked alongside their civilian counterparts, and ultimately made up about half the hand-computing operation. Several of the original townsite volunteers, such as Betty Inglis, Mici Teller, and Beatrice Langer, remained in the group through the end of the war.[16, 24]

A growing personnel roster was not unique to the hand-computing operation. The Lab's early leadership had severely underestimated the scope of the undertaking at Los Alamos, and initially believed that only about 300 research and support personnel would be needed on the Hill. By the end of the war, approximately 7,000 were living and working at Los Alamos. As the size of the operation ballooned, so did the hand-computing group, ultimately growing to 25 computers, making it the largest group in T Division.[25, 24, 21, 22]

T-Division leader Hans Bethe recruited the mathematician Don Flanders as one of the late-summer arrivals. An expert in transfinite numbers, Flanders became the first formal leader of the hand-computing operation, and the T-5 leader when the division established a group structure. He soon instituted a collection of changes to the operation. Computing personnel were previously allowed to choose from the three calculator models Frankel and Nelson had purchased. For the sake of computational speed and repair simplicity, Flanders standardized the operation around the fast Marchant Silent Speed line of 10-digit calculators. T Division had already abandoned all of the far less productive 8-digit machines the Lab had procured in the Spring. Flanders dispensed with the Monroe calculators entirely, which lacked multiply and divide functions (requiring operators to repeat additions or subtractions), and eliminated most of the Fridens, except for the two that Betty Inglis and Mary Frankel insisted on keeping. Flanders also instituted special columnar forms that he designed to speed up computation times.[13, 17, 10]

While Flanders streamlined and formalized the once ad-hoc operation, he also worked to maintain staff morale. He used his song-and-dance talents to write and perform in a comedy ballet that poked fun at the strangeness of life in wartime Los Alamos. Flanders himself played the part of a dancing General Groves, alongside a noisy mechanical-calculator prop that produced comically inaccurate calculations.[16]

Despite the expansion of the hand-computing group over the summer, and Don Flanders' improvements to its operation, in late 1943, the computing group struggled to produce reasonably accurate and timely neutron-diffusion calculations for the gun-type weapon. T

Division would ultimately not devise solutions to the neutron-diffusion problems in a form that the hand-computing group could process quickly and efficiently until the summer of 1944. Until then, the odd shape of the gun-type weapon's assembly made it extremely complicated to calculate estimates of the number of critical masses in the completed assembly, likewise with the problem of how to distribute safely the active fissile material between the two halves of the weapon's subcritical masses. The variety of ad hoc methods initially used at Los Alamos to calculate these problems pushed traditional computing methods to their practical limits, but the Lab could not adequately scale the hand-computing operation to meet the demand, not with housing and other resources on the Hill being at a premium. Non-traditional methods were needed if Los Alamos were to complete its mission under the time pressures of war. For discussions on the advancements in algorithms and computation that had occurred at Los Alamos by the summer of 1944, see Ferguson and Hill's accompanying article on neutronics algorithms, and the multiple articles on the Bethe-Feynman equation, in this issue.[11, 8, 4, 26, 27]

**Mechanizing the Operation**

Stanley Frankel and Eldred Nelson brought the computation problem to Hans Bethe in October of 1943. Bethe, in turn, raised the issue with Los Alamos' Governing Board, the collection of division leaders, administrative officers, and others who provided weekly counsel to Oppenheimer. Dana Mitchell, the Lab's procurement officer and a member of the Governing Board, offered a possible solution.[8, 6, 11]

During the year before his arrival at Los Alamos, Mitchell was in charge of procurement for the physics department at Columbia University, which housed the astronomer Wallace J. Eckert's Astronomical Computing Bureau. The Bureau was located in the same building where Enrico Fermi, Leo Szilard, and others had worked to produce a self-sustaining neutron chain reaction, before being moved to Chicago after Pearl Harbor. Eckert, an influential numerical astronomer who later provided lunar-orbit calculations for the Apollo missions, required a large number of differential equations in the 1930s for his work on the motions of the moon. This computational effort was tedious and time-consuming for human computers, but Eckert was aware of the Columbia Statistical Bureau's use of IBM punched-card accounting machines (PCAM) in its operation, and he believed the machines to be compatible with his requirements. In 1929, the astronomer Leslie Comrie at the British Nautical Almanac Office had launched the first effort to calculate lunar tables using punched-card equipment, and Eckert hoped to expand upon this work in the United States.[11, 28, 8, 30, 31]

Eckert persuaded IBM CEO Thomas Watson, Sr. to equip Columbia's new Astronomical Computing Bureau, later renamed after Watson. Based on Eckert's guidance, IBM modified the donated equipment for scientific use, and the company agreed to sell the modified machines to other astronomers in the US. Previously focused entirely on business and data processing, Eckert's close partnership with IBM formally introduced the company to scientific computation as a viable market for its products. This began a long relationship between Big Blue and the scientific community that would flourish in the early electronic-computing era. By the early 1940s, when Los Alamos began its search for new computing capabilities, the PCAM operation at Columbia had matured into one of the most sophisticated mechanized-computing facilities in the world. Eckert himself authored the first book detailing the use of accounting machines in science, published in 1940. His influence was far-reaching. During World War II, over a dozen mechanized computing operations based on that at Columbia entered use in strategic organizations in the US, including at the Ballistics Research Lab and the one Eckert himself founded at the US Naval Observatory. Following the war, Eckert assisted IBM in its development of the electromechanical SSEC computer that Los Alamos famously used for early H-bomb calculations.[31, 32]

Hans Bethe, who had also witnessed Columbia's PCAM operation in action, agreed with Dana Mitchell's suggestion to purchase accounting machines for the computation problem. Mitchell believed that the machines had the potential to shorten the computation time of a single calculation from many months to a few weeks. As a result, the Lab's Governing Board ordered three IBM 601 Multiplying Punches and accompanying support equipment. Although Oppenheimer urged a rapid delivery of the machines to aid with time-sensitive calculations needed for the fabrication of critical components, the equipment would not arrive until the early spring of 1944. During the delay before the equipment could arrive, the research priorities at Los Alamos had begun to change, and computing assumed even greater importance to the success of the Lab's mission. For an in-depth discussion of the wartime computing facility, see Bill Archer's article on the topic in this issue.[11, 8, 33]

**Implosion**

The implosion concept received little attention at the 1942 Berkeley conference, but the physicist Seth Neddermeyer instigated a small-scale implosion experimental program during Los Alamos' first year. His experiments drew the attention of the Hungarian mathematician John von Neumann in the fall of 1943. An expert in hydrodynamics, von Neumann had already served as a wartime consultant for multiple scientific and military research centers in the US before Oppenheimer invited him to consult for Los Alamos. Examining Neddermeyer's implosion tests with metal cylinders, von Neumann argued that using large volumes of shaped conventional explosives to compress a sub-critical core might need less-costly fissile material, and less material in total, than the gun concept. As a consequence, the Lab's leadership began to take implosion more seriously by the start of 1944.[34, 8, 11]

This was a fortunate development for the Lab. By the summer of 1944, experimental results from Emilio Segre's group had revealed that a plutonium-based gun-type weapon would not work. The weapon could not assemble quickly enough to prevent predetonation—the premature initiation of fission—caused by the spontaneous fission of plutonium 240. The problem was that the Manhattan Project would likely have enough uranium for only a single bomb by the summer of 1945, but enough plutonium for multiple weapons. The Lab had little choice but to embrace implosion to utilize the available plutonium.[4, 8, 7]

In the summer of 1944, Los Alamos proceeded to "throw the book" at the implosion problem, but the computational work on implosion had begun much earlier. Following John von Neumann's discovery in late 1943, and the Lab's subsequent, increasing interest in implosion, multiple schemes for computing implosion hydrodynamics were advanced for use at Los Alamos. Von Neumann proposed the first numerical method in late 1943, which he soon improved upon. In February 1944, Rudolf Peierls arrived at Los Alamos as part of the British Mission to the Manhattan Project, and T Division embraced the method Peierls suggested, which was based on his previous experience with airborne blast waves. Shortcomings with both von Neumann and Peierls' methods when dealing with strong shocks led von Neumann, Peierls, and the physicists Tony Skyrme (from the British Mission) and Robert Richtmyer to continue developing improved hydrodynamic numerical methods. For an in-depth discussion of wartime hydrodynamics at Los Alamos, see Morgan and Archer's article on Lagrangian hydrodynamic methods in this issue.[4, 35]

A large volume of hydrodynamics differential calculations were needed for the implosion weapon's development, but employing hand computing for the complex implosion calculations was not practical, particularly if a realistic equation of state were to be used. The IBM PCAM equipment offered a solution. Although the equipment had not yet arrived, T Division soon began to calculate the beginning conditions for integrating the implosion partial differential equations on the machines.[10, 34]

Performing an implosion simulation on the IBM equipment began with punching the initial conditions of the simulation onto batches of 80-column IBM punched cards, which operators fed through the machines in sequence. The implosion simulation process involved integrating coupled nonlinear differential equations through time steps. The cycle of a card deck through the equipment in the machine room represented the integration of a single time step. Each cycle required about a dozen separate steps on the machines, with human operators intervening often in the cycle.[10, 34, 14]

In March 1944, T Division needed to test the numerical process for the PCAM equipment, but the machines had not yet arrived. To save time, Frankel, Nelson, and Richard Feynman tested the process with the aid of the hand-computing group. A computer using a Marchant performed a single task in the cycle. One person would add, another would multiply, and so on, with each computer passing their results to the next using index cards. The process was efficient enough that the human computers during the test operated at speeds predicted for the PCAM equipment.[36]

The machines arrived before John Johnston, the previously drafted IBM technician who Los Alamos had requisitioned through the Army. Because of project secrecy, IBM could not send an installation crew. Feynman, Frankel, and Nelson took it upon themselves to begin assembling the equipment on their own, having only wiring blueprints as assembly instructions. They installed the machines one floor below the hand-computing group in the wooden building next to the pond. When Johnston arrived two days later, he discovered that some of the timing cams needed to be adjusted, but the machines were otherwise functional. Hans Bethe contended that Feynman, Frankel, and Nelson were the first outside IBM to assemble punched-card machines successfully. Johnston became a valued member of the IBM operation, continuing after the war. Wanting to test the assembled IBM equipment, in April, Feynman orchestrated a direct competition between the hand-computing group and the machines, resulting in a near tie for two days, until the machines only pulled ahead once the human computers began to fatigue from the rapid pace.[11, 14]

In early spring of 1944, the theoretical demands of implosion warranted a reorganization of T Division to address the problem. Bethe organized the division into a group structure more common in other divisions. He assigned Edward Teller as the first leader of the T-1 Implosion group, in charge of the mathematics of implosion, and calculating the time of assembly. Robert Serber's T-2 Diffusion Theory group initially housed the PCAM operation, before it was transferred to the new T-6 IBM Computations group later in 1944. Frankel and Nelson were T-6's first co-leaders. Teller's fascination with the Super soon distracted him from his responsibilities as leader of the implosion group. Bethe replaced him with Rudolf Peierls in July. Teller transferred to Enrico Fermi's F Division, where he continued his work on the Super.[8, 4, 24]

## Operating the Machines

Members of the Special Engineer Detachment (SED), army recruits who had scientific and technical backgrounds, were the primary operators of the Los Alamos IBM PCAM equipment during the war. The SEDs included skilled mechanics, machinists, graduate students, and other soldiers who demonstrated sought-after technical skills. They worked across many parts of the Manhattan Project, with over 475 SEDs at Los Alamos alone by the end of 1943. SEDs also provided some of the computational aid in T Division outside either of the computing groups. Peter Lax, for example, who won the Abel Prize in mathematics after the war, worked directly for Robert Serber, performing hydrodynamics calculations. One of the SEDs in the IBM group, John Kemeny, later gained notoriety as co-developer of the BASIC programming language, and eventually rose to become president of Dartmouth. Kemeny cited his experiences with the Los Alamos computing operation, and particularly Los Alamos consultant John von Neumann, with the driving fascination in mechanized computing that guided his career. The PCAM operator Harold Ninger and keypunch-operator Frances Noah were among the few civilians recruited for the group. Recruiting enough operators was only the first personnel challenge for the PCAM operation, the other was finding someone to train them.[10, 37, 38, 39]

## Rare Expertise

Naomi Livesay was one of a very small number of experts in the world in the use of IBM punched-card equipment for scientific applications when she arrived at Los Alamos. Although Livesay did not originally come to the Lab to work on the IBM equipment, her timely arrival and her rare skill set were critical for the PCAM operation's success. A brilliant mathematician, Livesay earned multiple degrees in the subject. She completed the coursework for a Ph.D. at the University of Wisconsin, but the Mathematics Department would only award her an education-oriented Ph.M., because her professors believed that women had no place in mathematics.[16, 8, 10]

Joe Hirschfelder, who later taught the computer training course at Los Alamos, taught chemistry at Wisconsin while Livesay was a graduate student. Believing the university faculty had treated her unfairly, Hirschfelder connected Livesay with a job at the Princeton Surveys, where she underwent six-months of training with IBM PCAM equipment. Livesay learned how to "program" the machines by rewiring their plugboards. This altered the machines' internal electrical connections, causing them to move and manipulate punched-card data in a variety of ways. Her work won her a prestigious Rockefeller Foundation Fellowship at the University of Chicago. When a Chicago faculty member arranged for the fellowship not to be renewed, Livesay became an instructor at the University of Illinois, where, in late 1943, she received a letter from Joe Hirschfelder with a mysterious job offer.[16]

She arrived at Los Alamos in February 1944 only to discover that she no longer had a job. Joe Hirschfelder's group had been working on the ill-fated plutonium gun-type weapon, and the group had just been terminated. Intrigued by her experience with IBM equipment, Hans Bethe arranged an interview for Livesay with Stanley Frankel and Eldred Nelson. Initially hesitant about the job, and wary of an "odd character" who later introduced

himself as Richard Feynman, Livesay accepted the post, with the title of Assistant Scientist.[16]

The only person at Los Alamos who had formal training in the operation of IBM accounting machines, it was up to Livesay to design and implement the plugboard programs needed for the implosion simulations, although Stanley Frankel also became proficient in the process through self-training. A machine's plugboards often had to be swapped between steps of an equation to carry out new operations. Livesay also trained and supervised the crews of about twenty SEDs and civilians who operated the machines. Because of Livesay's timely arrival, implosion work started only about a week after the equipment arrived. The punched-card operation ran for three shifts, twenty-four hours per day, six days per week.[16, 8, 10, 11]

The first implosion simulations on the PCAM equipment explored different configurations of the implosion device and its components. These exploratory simulations were used to select a configuration for detailed modeling, which then informed the selection of a design for construction into a working device. Calculating the weapon's explosive efficiency and yield followed. An implosion simulation began with modeling the detonation of the weapon's high-explosive charge, then it followed the resulting shockwave as it propagated through the core and reflected back. The calculations had to be transferred to human computers when the shockwave encountered a boundary between two materials, due to the complexity of the interface calculations. Once the simulated shockwave returned to a uniform material, the calculations went back to the machines. Interface calculations were only one of the many points where operators had to intervene with the machines, locating errors was another.[8, 10, 16, 11]

The dust that caused excessive wear on the Marchants would cause any of the hundreds of electric relays in an accounting machine to stick, potentially causing errors. Metropolis later recalled that, partly because of this problem, the numerical procedure was deliberately designed to be "insensitive to small errors." In the interest of speed (there was no time to repeat the computation of whole problems), errors in the least significant digits were ignored, so the operators only had to correct errors in the more significant digits. Nevertheless, some parts of an integration step needed to be run through the machines several times.[16, 14]

The work began slowly in the spring of 1944. Although the results of the first implosion problem were considered "very satisfactory," and pointed toward the feasibility of an implosion device, it had taken three months for the SEDs to complete. Machine errors were only part of the slow progress. Feynman believed a significant part of the issue was in keeping the SEDs uninformed about the purpose of the project. After obtaining permission from Oppenheimer, Feynman told the SEDs about the goals of the Manhattan Project, and their important role in fighting the war. No similar courtesy was extended to the hand-computing group.[11, 16]

Feynman argued that being told of the Lab's mission invigorated the SEDs, who quickly began to develop their own programs to improve the speed of the operation. An important improvement the SEDs devised was to color-code the cards governing each cycle. Previously, only a single problem was processed at a time, with the cards from that problem cycling through the equipment in the machine room. This left some of the machines sitting idle while waiting for a new cycle of cards. Color coding the cards allowed the SEDs to run two or three problems simultaneously, just out of phase with one another. This alteration played a significant part in the reduction of the needed processing time of a problem from three months to about three weeks.[16]

To meet the increasing computational demand, Naomi Livesay recruited Eleanor Ewing, a former colleague from the University of Illinois, as an assistant. With a graduate mathematics degree, Ewing also had extensive experience in physics, despite the Illinois faculty forcing her to sit in a separate, isolated row from the men in her physics classes. In 1943, she taught mathematics at Pratt and Whitney in Connecticut, but could not leave the tedious job because it was deemed essential to the war. To her surprise, in August 1944, Naomi Livesay offered her a new job on an unknown project in New Mexico.[16]

Ewing aided Livesay in programming the PCAM equipment and keeping it operating. The two shared an office, which John von Neumann also shared when he visited Los Alamos. After having visited Comrie's tabulating operation at the British Nautical Almanac Office, the same operation that inspired Wallace Eckert, von Neumann came to Los Alamos already interested in mechanized computing. Livesay and Ewing turned that interest into first-hand experience, showing the mathematician how to operate the PCAM equipment, and how to rewire tabulator plugboards. Although fascinated with the machines' capabilities, von Neumann's frustrating experiences with timing parallel

operations soured him on the concept of parallel computations in electronic computers.[16, 34, 8]

Von Neumann's exposure to mechanized computing at Los Alamos was formative for his interest in the potential of computing machines to overcome the limitations of traditional methods in science and mathematics. Until his untimely death in 1957, von Neumann served as one of the most influential figures in the development of electronic computing, and in the application of computing to problems in science.[34]

The Lab's wartime need for more and better computing resources sent von Neumann and other Los Alamos representatives across the US in search of new machines and capabilities. In the spring of 1944, Eldred Nelson negotiated with IBM vice president John McPherson to provide Los Alamos with a special line of triple-product multipliers, modified versions of the standard 601 Multiplying Punch, which the company also produced for several other sites in the US. The two also negotiated for the construction of a completely unique version of the 601 that would be capable of division. Dividing on the existing equipment was a time-consuming process of matching a card in need of division with a table of reciprocals on a card deck, then multiplying the reciprocal by the number being divided. IBM agreed to produce the requested machines.[10, 8, 40]

Arriving toward the end of 1944, two of the four triple-product multipliers that the Lab received were capable of division, and were unique to Los Alamos. Replacing the previous 601 Multiplying Punches, the new equipment improved the pace of implosion simulations by reducing the number of manual operations in an integration step. Los Alamos also sought IBM's aid in locating additional mechanized-computing labs that could augment the space- and personnel-limited operation at Los Alamos. IBM connected Los Alamos with Wallace Eckert, whose PCAM operation at Columbia inspired the Lab's own computing facility. With Eckert's Astronomical Computing Bureau already fully booked for other wartime tasks, Thomas Watson, Sr. and John McPherson agreed to supply Columbia with an expanded collection of machines specifically to aid Los Alamos' classified work. Von Neumann and Richard Feynman each traveled to Columbia to aid in setting up the calculations. The classified results were mailed to Los Alamos without labels (making them meaningless to an uninformed observer), while Feynman carried the labeled master copy back with him. Even before the war's conclusion, IBM recognized Los Alamos as a major driving force in scientific computing, and sought to accommodate its extreme technical requirements. These ties between Los Alamos and IBM only strengthened after the war, as the company began to take scientific computing more seriously as a market.[8, 41]

**The Race to Trinity and Its Lasting Influence**

After the first IBM machines arrived at Los Alamos in early 1944, Stanley Frankel became fascinated and proficient with the capabilities of the equipment. Having additional expertise in mechanized computing at hand aided the Lab to an extent, but Frankel's fascination developed into what Richard Feynman called "the computing disease," where he became more interested in making the machines do interesting things than in working on the task at hand. This issue, and Frankel's reported difficulty in working harmoniously with the SEDs, culminated with his removal as co-leader of the PCAM group, and his reassignment to Edward Teller's group working on the Super. Eldred Nelson became the sole group leader, while Bethe assigned Nicholas Metropolis as deputy group leader. Richard Feynman, who worked well with the SEDs, served as a T-6 consultant, while also leading the T-4 Diffusion Problems group.[16, 11, 36, 24]

SEDs who experienced the application of computing to implosion problems at Los Alamos, such as Peter Lax and John Kemeny, were not the only members of T-6 to become influential figures in electronic-computing development and its applications in the decades after the war. As the pressure to complete the implosion calculations increased in late 1944, T Division recruited the mathematician Richard Hamming to help maintain and program the IBM equipment. Hamming later joked that he had been a "computer janitor" at Los Alamos, keeping the machines running so the physicists could get back to work. Wanda Hamming later joined Richard in Los Alamos, and worked as a computer for Enrico Fermi and Edward Teller. Hamming had known nothing about computing machines before he arrived at Los Alamos, but the PCAM operation made him realize that "experiments that were impossible in the laboratory were going to be possible with computers." In his estimation, computing was about to change science dramatically, and he wanted to play a part in that change.[39]

Remaining in Los Alamos for six months after the war to help compile a technical description of the Lab's punched-card operation, Hamming then went to Bell Labs. Sharing an office with Claude Shannon, often called "the father of information theory," Hamming was involved in nearly all of Bell Labs' most prominent achievements in computer engineering over the next fifteen years, including the error-correcting "Hamming" codes that bear his name today.[39, 43]

As John von Neumann searched for new developments in computing between 1944 and 1945, he conducted a series of invited lectures at Los Alamos, discussing the implications of new computing technologies and methods for the future of scientific research. Metropolis and others argued that von Neumann inspired in T Division an enthusiasm for large-scale computers and for mechanizing weapons calculations, which continued long after the war. In early 1945, von Neumann brought news of the ENIAC project, the first fully electronic, general-purpose computer, which he discovered after a chance encounter with one of its co-developers, Herman Goldstine. The technical approach that the ENIAC represented had the potential to improve computing speeds thousands of times over conventional methods. A calculation that a human computer needed 20 hours to complete, the ENIAC could perform in 30 seconds. However, the ENIAC would not be complete until the end of 1945, making it of little use for the immediate implosion problem. Edward Teller, who worked closely and well with von Neumann, took notice of the ENIAC's potential for his group's work on the Super. Von Neumann arranged for Los Alamos to run the first large-scale program ever, a thermonuclear feasibility study, on the ENIAC, and Stanley Frankel, having been moved to Teller's group, was one of those who programmed the problem.[34, 23, 10, 13]

Frankel's first-hand experience with the ENIAC stoked his "computing disease" that began with the PCAM equipment. Over the following decades, both Frankel and Eldred Nelson emerged as influential figures in computing development on the West Coast of the United States, continuing even after the McCarthy era ended Frankel's government work. On projects ranging in scope from mainframes to early electronic desktop calculators, the two worked together and separately for academic, commercial, and defense organizations on numerous computing theory, design, and implementation efforts through the 1980s.[8, 41, 44, 45]

The "Los Alamos problem," filling one-million punched cards, ran on the ENIAC in December of 1945, after the war had ended. Von Neumann became a consultant on the ENIAC project, and worked with the project team on the design of the ENIAC's successor, helping to formalize the concept of a stored-program computer. Herman Goldstine later joined von Neumann at the Princeton Institute for Advanced Study (IAS) as the assistant director of the project that built von Neumann's IAS computer. Nicholas Metropolis based the design of Los Alamos' MANIAC on the IAS machine, as did IBM with its first fully electronic computer, the 701 "Defense Calculator." Unveiled by Oppenheimer, the 701 was first demonstrated to the public running an uclassified program from Los Alamos, and the Lab received the first 701 installed outside IBM.[13, 11, 46, 28]

## Computing and the Selection of Trinity

After its slow start in the spring of 1944, the punched-card group, with intervention from the human computers when needed, had finished eight implosion problems by the end of 1944, and seventeen in 1945. Those simulations aided in ruling out the initial design that project scientists had focused on until late 1944. Modeling revealed that implosion asymmetries with that design's hollow plutonium sphere might have prevented assembly. Exploratory simulations of a solid-core design first proposed by theoretical physicist Robert Christy in September 1944 were more encouraging. As a result, T Division and the Lab's leaders chose to run detailed simulations of the "Christy Gadget's" expected behavior. For a detailed examination of the contributions of Robert Christy, Rudolph Peierls, John von Neumann, and others to the "Christy Gadget," see T. A. Chadwick and M. B. Chadwick's accompanying article in this issue. Simulations revealed that the Fat Man-type weapon would produce a good energy yield, and was the most practical path toward an atomic device. As a consequence, the Lab's leadership chose that design for construction. The Trinity test in July of 1945 verified the calculations for the Fat Man design.[8, 10, 47]

## Conclusion

The computing effort at wartime Los Alamos, human and mechanized, has received fragmentary examination in the available historical literature, often obscuring the impact of computing on the success of the Trinity test and the completion of the Lab's mission. Computing was not the sole cause of Trinity's success, but the simulation data provided material aid to the design and development choices made by the Lab's scientific and engineering staff for what became the Trinity device, along with the multitude of other decisions that numerical methods helped to guide at Los Alamos throughout the war. Mechanizing the computing effort with accounting machinery greatly accelerated the implosion calculations, and helped to complete those problems amid the breakneck pace of the Manhattan Project. This computing effort also had a lasting impact at and beyond Los Alamos in the decades that followed.[8]

A vast collection of technical and human networks converged at Los Alamos to find answers for the largest scientific R&D effort in history. People from a variety of backgrounds used their skill and ingenuity to perform hand calculations and to adapt primitive accounting machines to provide critical information that helped guide that R&D effort to success. The experiences and inspiration they carried with them beyond the wartime Lab, and the fields of scientific and technical research they created or revolutionized as a result, also made Los Alamos a key driver in scientific-computing development through to the present.

**Acknowledgements**

My thanks to Bill Archer for his valuable insights and input into the history of Los Alamos computing and weapon physics. I would also like to thank the reviewers, in addition to Bill, whose comments made this a better article.

This work was supported by the US Department of Energy through the Los Alamos National Laboratory. Los Alamos National Laboratory is operated by Triad National Security, LLC, for the National Nuclear Security Administration of the US Department of Energy under Contract No. 89233218CNA000001.

[33] B. J. ARCHER, "The Los Alamos Computing Facility During the Manhattan Project," *Weapons Review Letters*, this issue (2021).